# Introduction of water-vapor broadening coefficients and their temperature dependence exponents into the HITRAN database, Part I: $CO_2$, $N_2O$, CO, $CH_4$, $O_2$, $NH_3$, and $H_2S$

**Y. Tan[1,2], R.V. Kochanov[1,3], L.S. Rothman[1], I.E. Gordon[1]\***

[1] Harvard-Smithsonian Center for Astrophysics, Atomic and Molecular Physics Division, Cambridge MA, USA.

[2] Hefei National Laboratory for Physical Sciences at Microscale, iChem center, University of Science and Technology of China, Hefei, China.

[3] Tomsk State University, Laboratory of Quantum Mechanics of Molecules and Radiative Processes, Tomsk, RUSSIA.

Corresponding author: Yan Tan (tanyan@ustc.edu.cn) ; I.E. Gordon (igordon@cfa.harvard.edu)

**Key Points:**

- Water-vapor broadening parameters for molecules in the HITRAN database are introduced for the first time
- $H_2O$-broadening effects for spectral lines of different molecules are discussed
- Procedures describing how to work with new parameters using HITRAN*online* and HAPI are explained.

**Abstract**

The amount of water vapor in the terrestrial atmosphere is highly variable both spatially and temporally. In the tropics it sometimes constitutes 4-5% of the atmosphere. At the same time collisional broadening of spectral lines by water vapor is much larger than that by nitrogen and oxygen. Therefore, in order to accurately characterize and model spectra of the atmospheres with significant amounts of water vapor, the line-shape parameters for spectral lines broadened by water vapor are required. In this work, the line-broadening coefficients (and their temperature dependence exponents) due to the pressure of water vapor for lines of $CO_2$, $N_2O$, $CO$, $CH_4$, $O_2$, $NH_3$, and $H_2S$ from both experimental and theoretical studies were collected and carefully reviewed. A set of semi-empirical models based on these collected data was created and then used to estimate water broadening and its temperature dependence for all transitions of selected molecules in the HITRAN2016 database.

**1 Introduction**

The current edition of HITRAN2016 (Gordon et al., 2017) has substantially increased the potential for the database to model radiative processes in terrestrial and planetary atmospheres. The previous editions of the database (from HITRAN2004 (Rothman et al., 2005) to HITRAN2012 (Rothman et al., 2013)) provided a limited set of parameters for each line transition, fixed within a 160-character record in ASCII files. Within that parametrization, only self- and air-broadening parameters, temperature dependence of the air-broadening parameters for the line list of each HITRAN molecule were available. The new edition has introduced line broadening parameters due to pressure of $H_2$, He and $CO_2$ for molecules of planetary interest including $SO_2$, $NH_3$, HF, HCl, OCS and $C_2H_2$ for the first time (Wilzewski et al., 2016). Later on the corresponding data for the CO molecule broadened by planetary gases was added from Li *et al.* (Li et al., 2015). This has instigated a significant progress for generating high-precision molecular absorption cross-sections relevant to the studies of planetary atmospheres. This new initiative took full advantage of the new structure of the HITRAN database (see (Hill et al., 2016)) which allows storage and effective retrieval of these parameters. The HITRAN Application Programming Interface (HAPI) (Kochanov et al., 2016) also makes good use of these new parameters allowing the calculation of cross-sections at different proportions of ambient gases (see Figure 31 of the HITRAN2016 paper (Gordon et al., 2017) for instance).

In this work we continue to build on this success and add line-broadening parameters due to the pressure of water vapor for $CO_2$, $N_2O$, $CO$, $CH_4$, $O_2$, $NH_3$, and $H_2S$.

It is well known that water vapor is highly variable in Earth's atmosphere, and it's also been confirmed in recent studies (Benneke & Seager, 2013; Hedges & Madhusudhan, 2016) that water vapor can represent a potentially significant cross-sensitivity source. Water vapor is a major absorber of the infrared light in the terrestrial atmosphere but it is also a very efficient broadener of spectral lines for other gases. The broadening by water vapor is much larger than that of nitrogen and oxygen which are the two main contributors to dry air broadening. Although nitrogen and oxygen are the most abundant terrestrial gases, water vapor does make an appreciable impact on the retrievals, especially in the tropics where water concentrations reach up to 5%. Fig.1 shows the simulated cross-sections for the $CO_2$ R(10) transition in the $\nu_3$ band at 2357.3207 cm$^{-1}$ and $CH_4$ transition at 6250.6943 cm$^{-1}$ broadened by air-, self-, and $H_2O$ using HAPI (Kochanov et al., 2016). For instance, satellite-based space projects including the NASA Orbiting Carbon Observatory re-flight(OCO-2) (Crisp, 2015), the TANSO-FTS on the Japanese Green-house Gases Observing satellite (GOSAT) (Kuze et al., 2009) and the Chinese TanSat satellite instrument (Chen et al., 2012) are proposed to retrieve surface pressure and column abundances of $CO_2$ with sub-percent precision. Central to the accuracy of remote-sensed $CO_2$ quantities, however, is the accuracy of the spectroscopic input (line positions, line intensities, and line-shape parameters) used in atmospheric models within the retrieval algorithms. The parameterization of the line shape includes pressure broadening and shift parameters, as well as their temperature-dependence exponents. Besides, they may also include some additional physical phenomena affecting line shape such as line-mixing, line narrowing, and speed-dependence. Since the ambient atmospheric surface pressure of water vapor could be up to 5% in the tropical regions, reducing uncertainties associated with water vapor is imperative to achieve a sub-percent precision in high-precision remote sensing for $CO_2$ retrieval. Interestingly, the atmospheres of rocky planets that may have suffered large impacts are expected to have "steamy" atmospheres (Benneke & Seager, 2013), and therefore knowledge of broadening of spectral lines by water vapor is important for modeling spectra of these atmospheres. In their recent paper, (Gharib-Nezhad & Line, 2019), emphasized the importance of using proper water broadening parameters when modeling the emission and transmission spectra as well as on the vertical energy balance in sub-Neptune/super-Earth atmospheres.

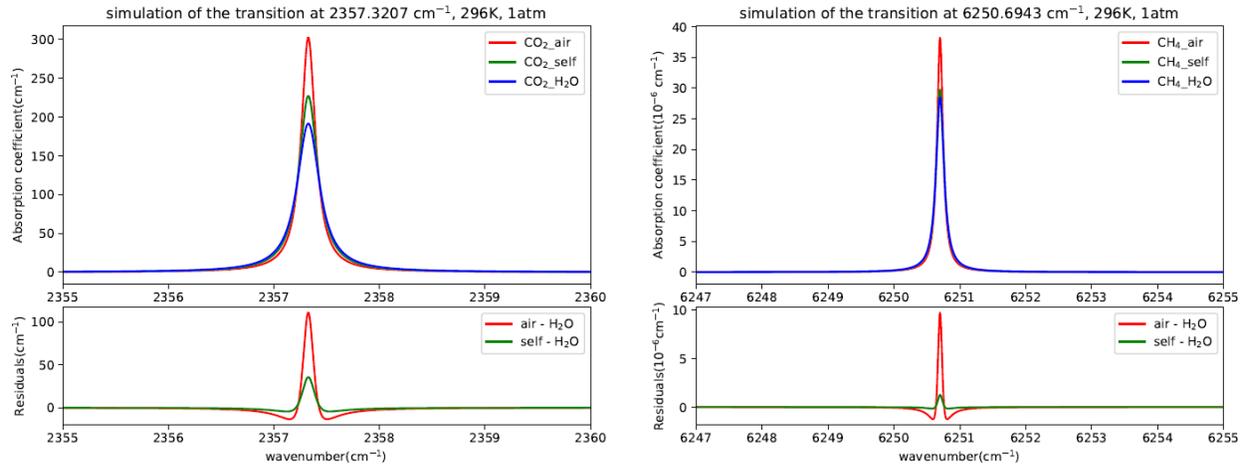

**Figure 1**. Comparison of simulated absorption cross-sections for the $CO_2$ transition at 2357.3207 cm$^{-1}$ and the $CH_4$ transition at 6250.6943 cm$^{-1}$ broadened by air, self, and $H_2O$. The lower panel represents the difference in absorption coefficients calculated with air- and self- broadening to that with water-vapor broadening.

The water-vapor broadening coefficients as well as their temperature dependence exponents have been assembled from both experimental and theoretical studies. The collected data were used to create semi-empirical models so we could populate the entire line lists of relevant molecules in the HITRAN2016 database with relevant parameters. Considering that the uses of these parameters may go beyond traditional HITRAN applications and could be used to model combustion processes or "hot" planetary atmospheres, we want these models to be applicable for the future editions of the high temperature database analog to HITRAN, HITEMP (Rothman et al., 2010) as well. Therefore, wherever there was enough experimental and theoretical data available, the models were based on the Padé approximants which allow smooth extrapolation to the high-rotational levels rather than using polynomials or fixing the widths of the high-*J* lines to the value of the last measured transition which was the case in the previous efforts (for instance in (Wilzewski et al., 2016) and (Li et al., 2015)). Associated programs in Python that allow calculating the water-vapor broadening parameters for $CO_2$, CO and $O_2$ even at very high *J*'s are provided in the supplementary section. The semi-empirical models derived here can be applied to all isotopologues of the molecules in question.

**2 Generated Datasets**

The line-shape parameters derived in this paper still follow the traditional HITRAN ".par" formalism where Lorentzian half widths (at $T_{ref}$ = 296K) are provided based on the Voigt profile measurements. The sophisticated line shapes beyond the Voigt profile were not considered in this work since the majority of the water broadening parameters that came from measurements and calculations in the literature was still based on the Voigt profile. The line broadening coefficients by water vapor and the temperature dependent exponents can be retrieved from HITRAN*online* (www.hitran.org) in user-defined formats (see section 4 for details). The conversion of the half width from a reference temperature ($T_{ref}$) can be described by the following power law:

$$\gamma(T) = \gamma(T_{ref}) \cdot \left(\frac{T_{ref}}{T}\right)^n$$

(1)

Here *n* is the temperature-dependent exponent which could also be determined from the slopes of the least-square fits of $-\ln \gamma(T)$ vs. $\ln T$. Then it leads to:

$$n_{H_2O} = -\frac{\ln \gamma_{H_2O}(T_{ref}) - \ln \gamma_{H_2O}(T)}{\ln T_{ref} - \ln T}$$

(2)

It is also worth emphasizing that the temperature dependence exponents derived from the power law above work only within a relatively narrow temperature regime. A recent study also shows that a double power law developed by Gamache *et al.* can model the temperature dependence of the half widths over large temperature ranges (Gamache & Vispoel, 2018). However, the data available for this fitting procedure are quite limited, and consequently we have chosen one temperature exponent at 296K in this work.

2.1 $CO_2$

After water vapor, carbon dioxide is the second strongest absorber of infrared radiation (including thermal IR) despite its low concentration in the Earth's atmosphere. This makes it an important greenhouse gas which is monitored by many ground-based and satellite-based instruments, for instance, the NASA Orbiting Carbon Observatory re-flight (OCO-2) (Crisp, 2015), the Greenhouse Gases Observing Satellite (GOSAT) (Kuze et al., 2009) and the TanSat satellite (Chen et al., 2012). The OCO-2 mission was proposed to reach the sub-percent precision of $CO_2$ concentration measurement which places extraordinary demand on the quality of spectroscopic parameters of $CO_2$ lines. Since water vapor is highly variable in the tropical regions, the pressure broadening by water vapor could make an additional contribution to the measured widths of $CO_2$ lines. It is thus important to introduce accurate half widths as well as temperature dependence exponents of the half widths to the HITRAN database in order to model $CO_2$ absorption cross sections for high accurate remote-sensing tasks to avoid the spectroscopic uncertainties associated with ambient water vapor.

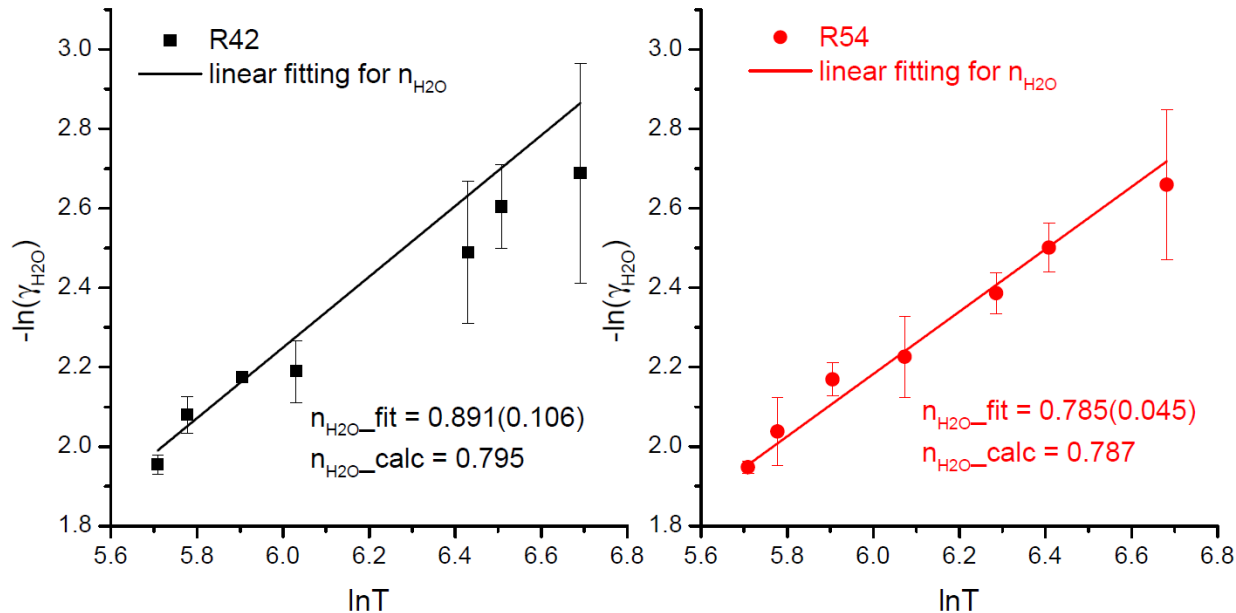

**Figure 2.** The temperature dependence exponents for the R(42) and R(54) transitions in the $\nu_3$ band of $CO_2$ determined from the slopes of the least-square fits of $-\ln \gamma(T)$ vs $\ln T$ from 300 to 805 K (Delahaye et al., 2016a; Rosenmann, Hartmann, et al., 1988; Rosenmann, Perrin, et al., 1988; Sung et al., 2009). The temperature dependence exponents derived from linear fitting agree well with the calculated parameters from (Rosenmann, Hartmann, et al., 1988).

A theoretical calculation for $CO_2$ broadening by water vapor has been made by Rosenmann *et al.* in 1988 (Rosenmann, Hartmann, et al., 1988). The semi-classical Robert-Bonamy formalism was used to predict the water-vapor broadening Lorentz half widths of $CO_2$ lines as well as their temperature-dependent exponents at a wide temperature range from 300 to 2400 K. It was demonstrated that the water-vapor broadening Lorentz half widths of $CO_2$ can reach twice the values of the air-broadening half widths. Until recently, direct measurements were very rare and existed for only two lines (R(42) and R(54) in the $\nu_3$ band) (Rosenmann, Perrin, et al., 1988). With increased requirements on the accuracy of atmospheric retrievals, more experimental data for $CO_2$ broadening by water vapor have emerged recently. The water-vapor broadening Lorentz half widths of 182 $^{12}CO_2$ lines in the $\nu_3$ and $\nu_2+\nu_3-\nu_2$ bands, as well as the $\nu_3$ band of $^{13}CO_2$, were reported by Sung *et al.* (Sung et al., 2009) in 2008 with a standard Voigt profile at near room temperature. There were three transitions of $CO_2$ broadened by water vapor near 1.57 μm that had been measured by speed-dependent Voigt profile at near room temperature (Wallace et al., 2011) which were not included in our fitting procedure because of the poor agreement with other measurements. More recently, (Delahaye et al., 2016a) have reported the transmission spectra of $CO_2$ in a high concentration of water vapor, also in the 4.3 μm region. The Lorentz half widths of 64 $CO_2$ transitions broadened by water vapor were determined at 323K and 367K (Delahaye et al., 2016a). All these theoretical and experimental results were used to generate the dataset for $CO_2$ transitions appropriate for the HITRAN database.

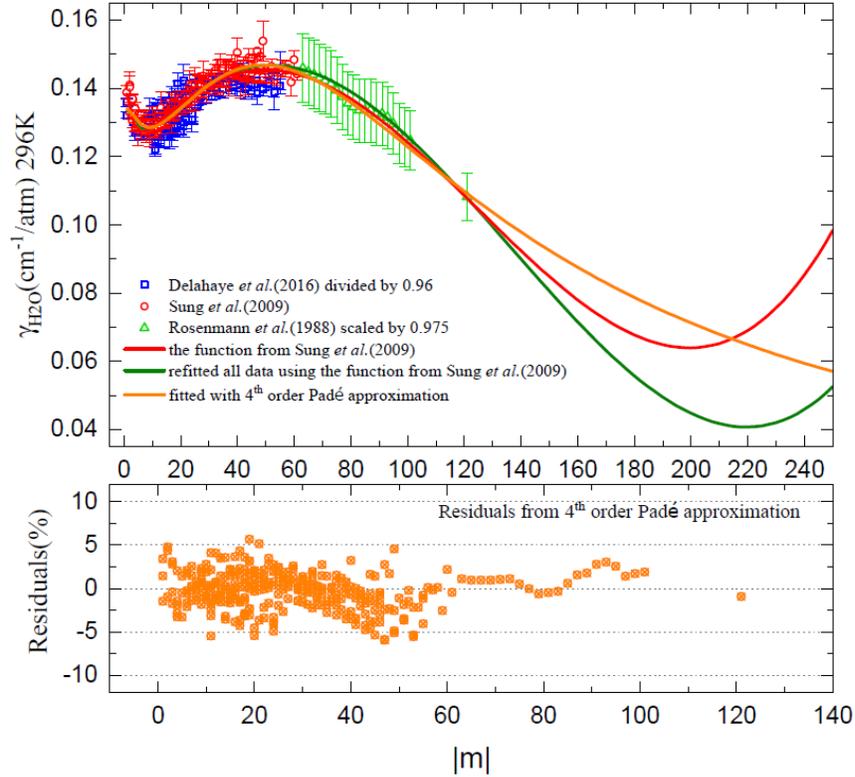

**Figure** 3. The upper panel shows the constructed water-vapor broadening Lorentz half widths for $CO_2$ transitions at 296 K based on the empirical fitting results from both experiments and calculations. The fitted half widths from different functions are shown here in solid lines of different colors which are valid up to $|m| \leq 121$. The lower panel shows the relative residuals fitted from the 3$^{rd}$-4$^{th}$ order Padé approximation with a standard deviation of 3%.

Based on the comparison of the temperature dependence exponents from the linear fitting of the power law and that of the semi-classical calculations, the theoretical predictions for the temperature dependences were used as shown in Fig. 2. The experimental results of $CO_2$ Lorentz half widths broadened by water vapor from (Delahaye et al., 2016a) at 323K and 367K, (Sung et al., 2009) at near room temperature 301K and (Wallace et al., 2011) at 294K were carefully converted to corresponding values at 296K by using temperature dependence exponents from the predictions of (Rosenmann, Hartmann, et al., 1988). Due to the large error bar for experimental results by Wallace *et al.* as well as the difference of the line profile used, these data were excluded from the final fitting procedure. There was about 4% unexplained systematic offset between the measurements of Delahaye *et al.* and those of Sung *et al.* in the same spectral region of 4.3 μm although they were sharing the same rotational dependence. (Sung et al., 2009) data is used for spectroscopic input for the OCO-2 mission (Oyafuso et al., 2017) and therefore for consistency, the experimental results converted from Delahaye *et al.* were then divided by 0.96 before fitting. In order to make more accurate predictions for transitions involving high-rotational quanta, the calculated half widths from (Rosenmann, Hartmann, et al., 1988) were also incorporated into the fitting (only for $63 \leq |m| \leq 101$ and m = 121) but scaled by multiplying by a factor of 0.975 to account for discrepancy between these calculations and measurements from (Sung et al., 2009). Here *m* is a running number which (when expressed as a function of lower-state rotational quantum number $J''$) is equal to $-J''$ for P-branch transitions and $J''+1$ for R-branch. (Sung et al., 2009) fitted their data to Eq. (3) and the set of coefficients recommended in their work is listed in Table 1. We also use the same function to refit all the data collected from experiments and calculations as shown in Fig. 3 and Table 1.

$$\gamma_{H2O}(|m|) = \frac{a}{m^2} + \frac{b}{|m|} + c + d \cdot |m| + e \cdot m^2 + f \cdot |m|^3$$

(3)

**Table 1.** Fitted coefficients (from Eqs. (3,4)) for calculating water-vapor-broadened Lorentz half-widths of $CO_2$ ( at 296 K and in the units of $cm^{-1} \cdot atm^{-1}$ ).

| Coefficients (Eq. (3)) | (Sung et al., 2009) | This work | Coefficients (Eq. (4)) | This work |
|---|---|---|---|---|
| $a$ | -0.03563 | -0.06539 | $a_0$ | -25.18047 |
| $b$ | 0.06318 | 0.09205 | $a_1$ | 459.98114 |
| $c$ | 0.1093 | 0.10547 | $a_2$ | 35.11405 |
| $d$ | 0.001498 | 0.00161 | $a_3$ | 0.67148 |
| $e$ | -1.848E-5 | -1.87E-05 | $b_1$ | 3141.5695 |
| $f$ | 4.924E-8 | 4.58E-08 | $b_2$ | 369.17545 |
|  |  |  | $b_3$ | -0.75135 |
|  |  |  | $b_4$ | 0.05441 |
| Valid range | $|m|\leq121$ | $|m|\leq121$ |  | $|m|\leq121$ |

In addition, the third-fourth order Padé approximant was also used to fit all data points as described here.

$$\gamma_{H2O}(|m|) = \frac{(a_0 + a_1 \cdot |m| + a_2 \cdot m^2 + a_3 \cdot |m|^3)}{(1 + b_1 \cdot |m| + b_2 \cdot m^2 + b_3 \cdot |m|^3 + b_4 \cdot m^4)}$$

(4)

The advantage of using this new function is that it extends standard Taylor treatments, overcoming possible convergence issues at high *m* values which could be very important at high temperatures. As shown in Fig. 3, when it comes to high *m* values the predictions from the polynomial function is increasing unphysically after $|m| = 180$ while the Padé approximant provides more reliable values. The fitting residuals scatters 3% between our fitted values and the experimental results.

In conclusion, the vibrational dependence of the water-vapor broadening half widths for $CO_2$ is almost negligible based on the available data, which is typical for linear molecules. Therefore, when estimating water-vapor broadening Lorentz half widths of $CO_2$ one should be concerned only about the rotational distribution which could be described from the 3rd-4th order Padé approximant as shown by Eq. 4, with coefficients from Table 1. The temperature dependence exponents for |$m$| less than 101 are taken from theoretical calculations by (Rosenmann, Hartmann, et al., 1988) and complemented with extrapolated values. The constant value 0.63 was used for |$m$| greater than 101.

### 2.2 $N_2O$

Nitrous oxide is one of the important air pollutants. Therefore the concentration of $N_2O$ in the Earth's atmosphere has also been monitored by different techniques in recent years. The water-vapor broadening parameters of $N_2O$ lines are needed for high-accuracy retrieval to determine their concentrations. Unfortunately, only one water-broadened transition of $N_2O$ has been measured (by (Deng et al., 2017)). Based on the ratio of the water-vapor broadening coefficients to the corresponding air-broadening coefficients, the scaling factor 1.9 is suggested to be used to multiply the air-broadening coefficients of $N_2O$ lines from the HITRAN database.

### 2.3 CO

The water-vapor broadening parameters of carbon monoxide lines are needed for the high-accuracy trace gas retrieval missions (Clerbaux et al., 2008; Liu et al., 2011, 2014), and these parameters are also very useful in the study of exoplanets with substantial amounts of water vapor in the atmosphere (Konopacky et al., 2013). The water-vapor broadening coefficients of CO transitions in the fundamental band have been measured by (Deng et al., 2017) and Willis *et al.* (Willis et al., 1984). (Henningsen et al., 1999) performed water-vapor broadening measurement for the R(7) transition in the 3-0 overtone band. The complete calculation for the water-vapor broadening parameters of CO lines in the 200-3000K temperature range had been presented by (Hartmann et al., 1988). They were made with a semi-classical model derived from the Robert and Bonamy (RB) approach.

As shown in Fig 4, the calculations were found to be systematically higher than those obtained from experiments by about 20%. Therefore, the calculated values were multiplied by the factor 0.81 according to our comparison with experimental results. The 3$^{rd}$-4$^{th}$ order Padé approximant was applied to fit all data from both theoretical values and experimental results while constraints were made at very high rotational levels to avoid negative results during the fitting. Therefore all data reported in ( Deng et al., 2017; Willis et al., 1984; Henningsen et al., 1999; Hartmann et al., 1988) were fitted with a least squares procedure to Padé approximant, see Eq. (4). The fitting results came with the coefficients: $a_0 = 8.72927$, $a_1 = -12.18068$, $a_2 = 15.04774$, $a_3 = -0.09195$, $b_1 = -0.57622$, $b_2 = 130.88058$, $b_3 = -0.71588$, $b_4 = 0.04076$. Then, the fitting results based on the Eq. (4) were used with |$m$| ≤ 150, and the HITRAN uncertainty code is increasing with rotational quantum number from 5 to 3 (see definition of HITRAN uncertainty codes in the documentation pull-down menu in HITRAN*online*, https//hitran.org/docs/uncertainties). For |$m$| > 150, a constant value 0.005 was used. The temperature-dependence exponents from the calculation in (Hartmann et al., 1988) were interpolated to cover the entire region with |$m$| ≤ 77. For |$m$| > 77, the constant value of 0.26 was used to apply to the rest of the lines.

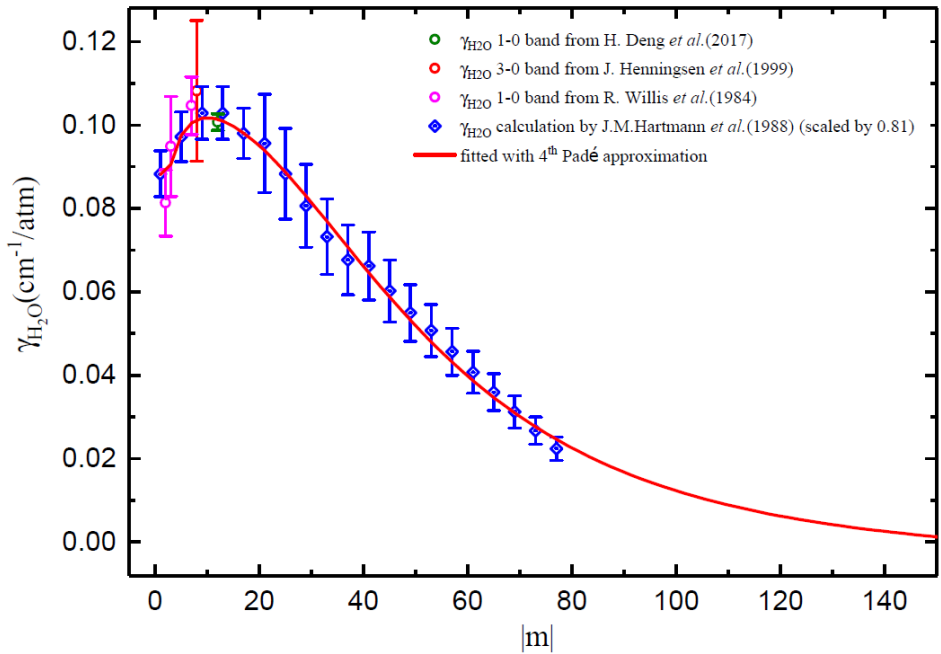

**Figure** 4. The pressure broadening coefficients of CO perturbed by water vapor. The green, pink and red points represent the broadening parameters for the R(11) transition in the fundamental band measured by (Deng et al., 2017); the P(2), P(3), P(7) transitions in the fundamental band are from (Willis et al., 1984); and the R(7) transition in the 0-3 overtone band by (Henningsen et al., 1999). The blue triangles were scaled calculations from (Hartmann et al., 1988) and the red line corresponds to the fitting result of the $3^{rd}$-$4^{th}$ order Padé approximant.

2.4 $CH_4$

Following water vapor and carbon dioxide, methane is also one of the most important greenhouse gases in the atmosphere and its warming potential is almost 23 times that of carbon dioxide over a one-hundred-year cycle (Solomon, S., D. Qin, M. Manning, Z. Chen, M. Marquis, K.B. Averyt & Miller, 2007). Therefore, the abundance of methane in the earth atmosphere is also one of the main objects in the remote-sensing missions including GOSAT (Kuze et al., 2009) and SCIAMACHY (Frankenberg et al., 2011), MERLIN (Ehret et al., 2017), CarbonSat (Buchwitz et al., 2013), and TROPOMI (Butz et al., 2012). For the purpose of reducing the uncertainties in the retrieval of atmospheric methane column amounts, it requires one to model the absorption cross sections of methane with an extremely high accuracy (to even sub-percent). It is therefore necessary to consider the contribution of the pressure broadening of $CH_4$ by water vapor besides the self- and air-broadening contributions (McDermitt et al., 2011; Miller et al., 2015). There were an appreciable amount of methane line shapes laboratory studies and theoretical calculations in recent years, but almost all of them were devoted to self- and/or air-broadening line-shape parameters of methane (see for instance (Devi et al., 2015, 2016; Ghysels et al., 2014; Hashemi et al., 2015; Smith et al., 2014)). The pressure-broadening parameters of $CH_4$ by water vapor in the mid- and near-infrared region were measured recently for the first time using a Fourier Transform spectrometer (Delahaye et al., 2016c). In that study 76 ro-vibrational transitions were measured. Before that, only the $2\nu_3$ R(3) and R(4) manifolds located at 1.6-μm region were studied from 316 to 580 K by diode laser absorption spectroscopy in (Gharavi & Buckley, 2005), and the ν$_4$ P(5), P(9) and P(10) manifolds located at the 8-μm region were measured by TDLAS at room temperature in (Lübken et al., 1991).

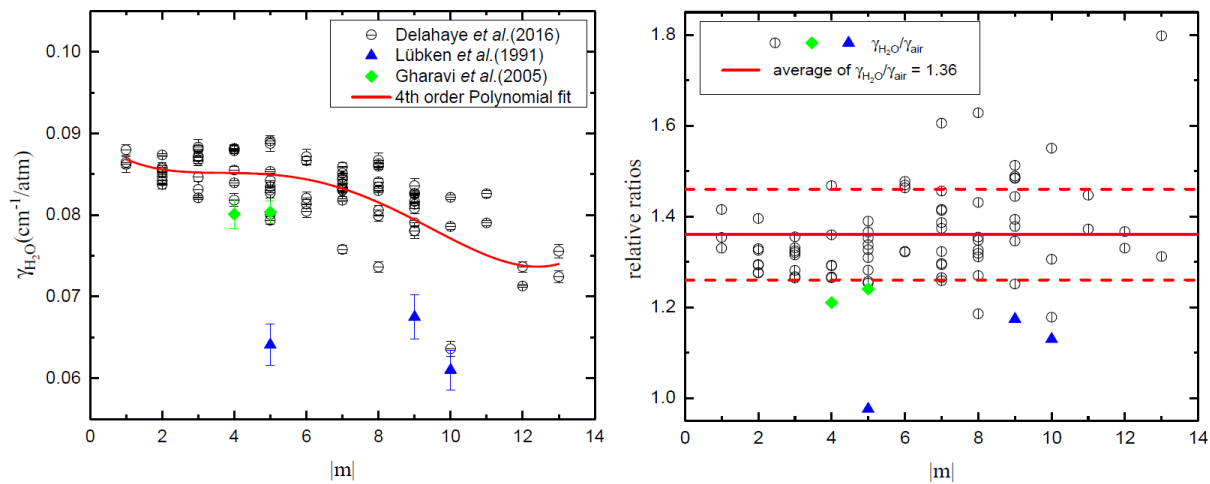

**Figure** 5. The pressure-broadening coefficients of methane perturbed by water vapor. (a) black circles represent the water-vapor broadening Lorentz half widths of CH$_4$ in the mid- and near-infrared region by (Delahaye et al., 2016c) with the red line for the 4$^{th}$ order polynomial fitting, and blue triangles from the ν$_4$ P(5), P(9) and P(10) manifolds by (Lübken et al., 1991), the 2ν$_3$ R(3) and R(4) manifolds in green diamond by (Gharavi & Buckley, 2005); (b) the ratio of the water-vapor broadening half widths of methane transitions to their corresponding air- broadening half widths in HITRAN, as well as the average value of the those ratios plotted here.

In general, the measured water-vapor broadening Lorentz half widths of methane were found to be about 20% larger than their air-broadening parameters as shown in Fig. 5. Since the rotational dependence of the water-vapor broadening effects for methane is still not quite clear based on our analysis of the available measurements, the water-vapor broadening coefficients at room temperature were then generated from the scaled value of their corresponding air-broadening parameters. The scaling factor of 1.36 (with a standard deviation of 0.10) was used based on the measurements from (Delahaye et al., 2016c), (Lübken et al., 1991) and (Gharavi & Buckley, 2005) to their corresponding air-broadening parameters in HITRAN. As for their temperature-dependence exponents, the scaling factor of 1.26 to the air-broadening temperature dependence exponents was used as well which also came from the experimental measurement of the 2ν$_3$ R(3) and R(4) manifolds in (Gharavi & Buckley, 2005) .

2.5 O$_2$

Molecular oxygen is the second most abundant gaseous component in the terrestrial atmosphere, and it is well-mixed in air. In many ground-based, airborne or satellite-based remote sensing projects (Blackwell et al., 2001; Cadeddu et al., 2007; Rosenkranz, 2001), the O$_2$ A-band at 0.76 μm is being used extensively to provide the information on atmospheric path length and surface pressure. It was also demonstrated recently that the 1.27 μm band can also be successfully employed in remote-sensing missions (Sun et al., 2018). Since the accuracy of the retrievals relies directly on the accuracy of the spectroscopic parameters from the input line shape information, the broadening parameters of the lines are of crucial importance to these remote-sensing missions. Meanwhile, water vapor is highly variable in the tropical regions; therefore it would be necessary to consider the pressure broadening of $O_2$ lines by water vapor in addition to the self- and air- broadening in order to reduce uncertainties in the spectroscopic input.

Unlike air- and self-broadening in the $O_2$ bands which have been studied extensively in recent years (Barnes & Hays, 2002; Drouin, 2007; Gordon et al., 2011; Long et al., 2010), only a few measurements have been performed for $O_2$ perturbed by water vapor. In 1994, (Fanjoux et al., 1994) presented the first extensive measurements of $O_2$-$H_2O$ broadening for the $O_2$ Raman Q-branch at 1553.3 cm$^{-1}$ in a wide temperature range (between 446 and 990K). Although the primary goal of that study was to support Raman thermometry of rocket engines, the authors had extrapolated their data to room temperature values based on the temperature dependent exponents fitted from the power law which would be comparable to other spectral regions. In recent years, a few more studies dedicated to water-vapor broadening parameters of $O_2$ lines in the pure rotational and A-band were made by using different techniques: the laser-based photoacoustic spectrometer (Vess et al., 2012), the frequency-multiplier spectrometer with a Zeeman-modulated absorption cell (Drouin et al., 2014), the radio-acoustic detected spectrometer (Koshelev et al., 2015), and Fourier transform spectrometer (Delahaye et al., 2016b). In order to provide accurate and reliable spectral parameters for atmospheric applications, the water-vapor broadening parameters of $O_2$ lines in the HITRAN database were then divided into two parts for generating the data set. As the first part, $H_2O$-broadening coefficients of $O_2$ lines in the A-band were obtained by multiplying $O_2$-air broadening coefficients in the A-band from the HITRAN2016 database by a single scaling factor of 1.1, as per recommendation by the most recent study from (Delahaye et al., 2016b). For the rest of the transitions, a complete analysis for all collected experimental data (presented in Fig. 6) was carried out. Note that the transitions with $N'' = 1$ were not included in our fitting procedure except for the pure rotational transitions from (Drouin et al., 2014) as shown in the grey-shaded area of Fig. 6. Their values fall off the pattern due to the large spin splitting in the lowest rotational level. Therefore the water-vapor broadening Lorentz half widths of $O_2$ for $N'' = 1$ were treated separately with the average value of all experimental results. The empirical function of (Drouin et al., 2014) (see Eq. 5) and the 3$^{rd}$-4$^{th}$ order Padé approximant were applied to fit all the other collected data with results listed in Table 2 for both functions.

$$\gamma(N'') = A_\gamma + \frac{B_\gamma}{(1 + c_1 \cdot N'' + c_2 \cdot N''^2 + c_3 \cdot N''^4)}$$

(5)

The temperature-dependence exponents for $H_2O$-broadened half widths of $O_2$ were fitted from data in (Fanjoux et al., 1994) and from Appendix A of (Drouin et al., 2014). For $N'' \leq 35$, the fitted and interpolated values used were based on the results of Table 4 from (Drouin et al., 2014), and the constant value 0.60 was used for the transitions with $N'' > 35$.

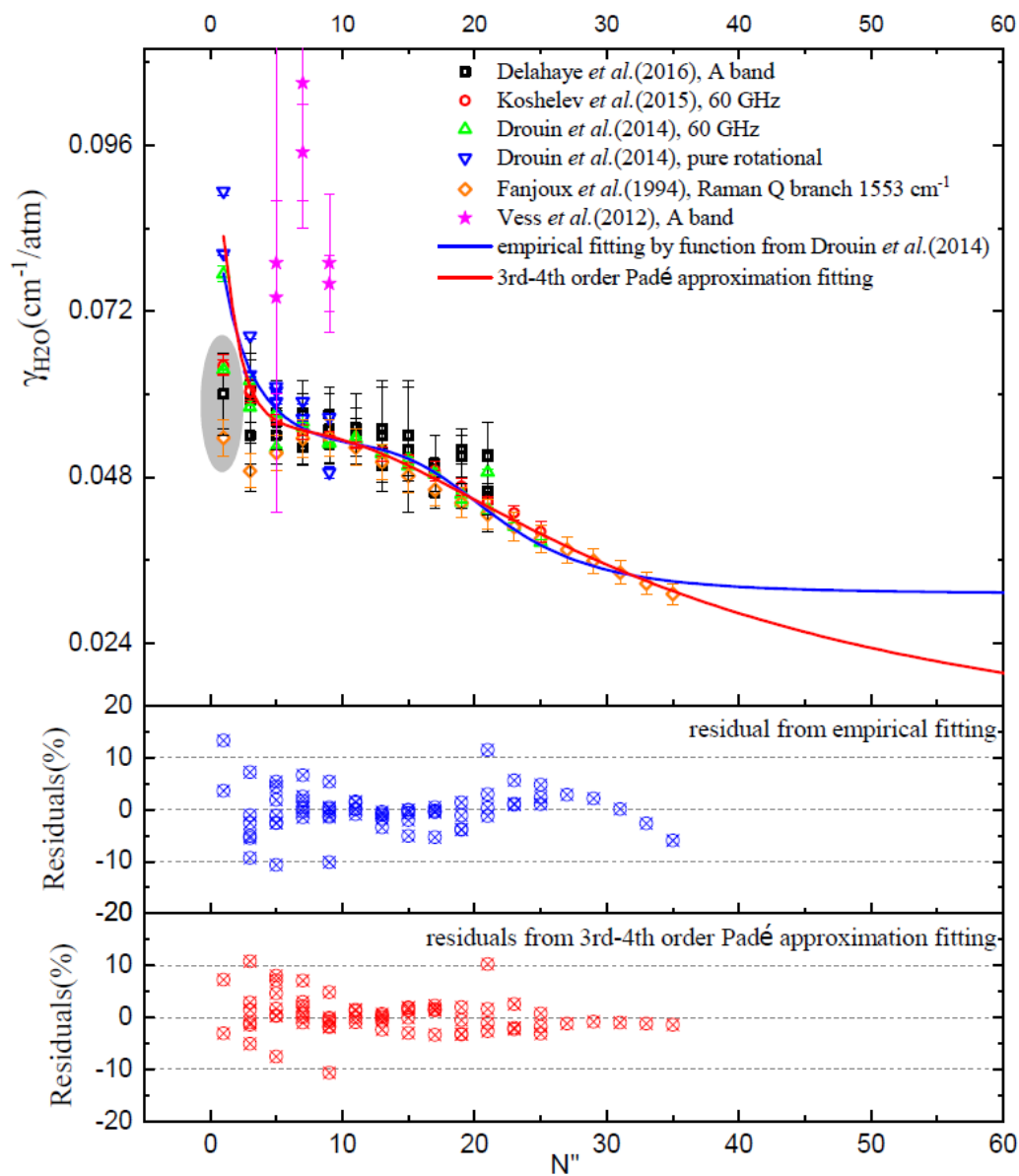

**Figure 6**. The upper panel shows the constructed water-vapor broadening Lorentz half widths for $O_2$ transitions at 296 K based on the fitting of the results from different experiments. The fitted half widths from different functions are shown here in solid lines of different colors which are valid up to $|m| = 35$. The lower panel shows the residuals fitted from the empirical function of (Drouin et al., 2014) and the $3^{rd}$-$4^{th}$ order Padé approximant with a standard deviation of 4% and 3%. Note here the grey-shaded area in the upper panel corresponds to the transitions with $N" = 1$ that were not included in our fitting procedure.

**Table 2.** Fitted coefficients (from Eqs. (4,5)) for calculating water-vapor broadened Lorentz half widths of $O_2$ ( at 296 K and in the units of $cm^{-1} \cdot atm^{-1}$ ).

| Coefficients (Eq. (5)) | (Drouin et al., 2014) | This work | Coefficients (Eq. (4)) | This work |
|---|---|---|---|---|
| $A_\gamma$ | 1.409 | 0.031 | $a_0$ | -5.19194 |
| $B_\gamma$ | 1.854 | 0.063 | $a_1$ | 7.43635 |
| $c_1$ | 0.382 | 0.383 | $a_2$ | 3.35387 |
| $c_2$ | -2.50E-2 | -2.30E-2 | $a_3$ | 0.40734 |
| $c_3$ | 4.06E-5 | 3.26E-5 | $b_1$ | -46.17 |
|  |  |  | $b_2$ | 116.8574 |
|  |  |  | $b_3$ | 0.43672 |
|  |  |  | $b_4$ | 0.35434 |
| Valid range | $|m| \leq 35$ | $|m| \leq 35$ |  | $|m| \leq 121$ |

Note: the transitions with $N'' = 1$ were not included in the fitting procedure and were treated separately in the final dataset generation.

### 2.6 $NH_3$

Ammonia is an important trace gas and is a subject of studies of different remote-sensing missions (Beer et al., 2008; Höpfner et al., 2016) . Its accurate spectroscopic parameters are needed for a variety of gas-sensing applications including those for environmental monitoring, industrial process control, and human breath analysis in medical science (Manne et al., 2006; Owen & Farooq, 2014; Stéphane Schilt et al., 2004). While each of these applications involved significant concentrations of water vapor in the system, the impact of water vapor on ammonia absorption features is not negligible for high-sensitivity detections of ammonia to ppb or ppm levels (Miller et al., 2015; Sun et al., 2015, 2017). In fact, recent experimental results from (S Schilt, 2010) had demonstrated that water vapor was a significant cross-sensitivity source especially in a high-temperature environment. However, not quite enough data about water-vapor broadening coefficients of $NH_3$ lines were available in the literature. The water-vapor broadening parameters of the $NH_3$ strong $v_2$ vibrational band around 1103.46 $cm^{-1}$ have been measured by (Owen et al., 2013). More recently, (Sur et al., 2016) also published the water-vapor broadening coefficients of $NH_3$ Q-branch transitions in the $v_2$ vibrational band. Therefore, there were fifteen transitions that had been studied in total as shown in Fig 7.

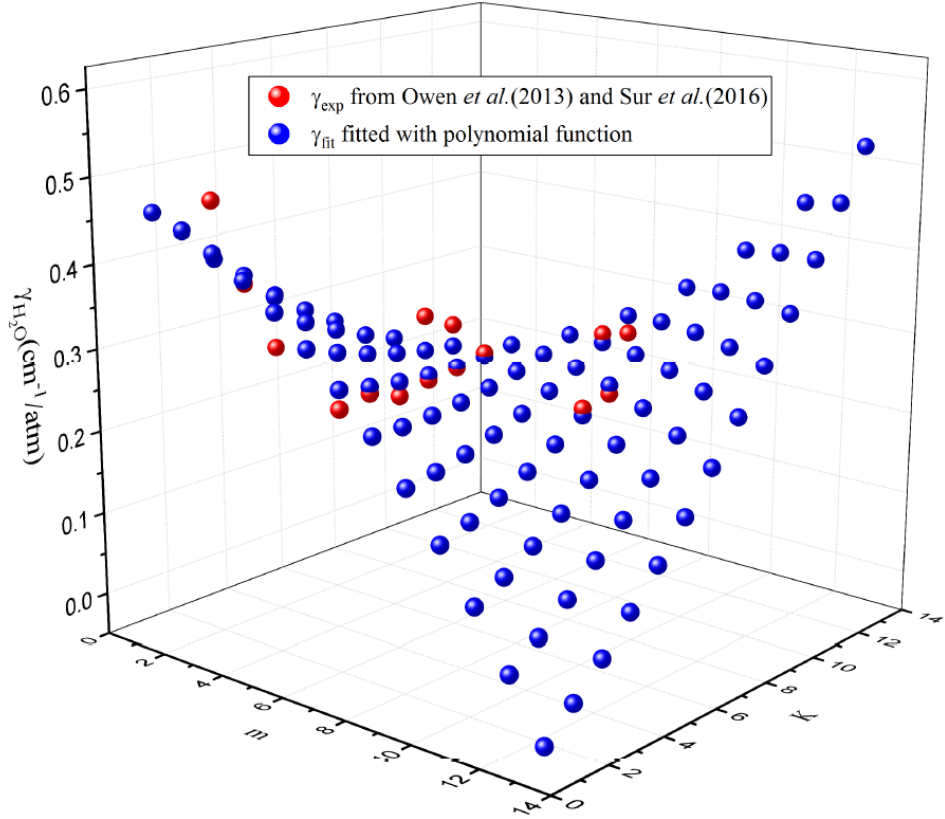

**Figure 7**. The pressure-broadening coefficients of $NH_3$ perturbed by water vapor versus the running index $m$ and quantum numbers $K$. The red points correspond to experimental values taken from (Owen et al., 2013; Sur et al., 2016) and the blue points correspond to the least squares fitting to the polynomial function Eq. (6).

The dependence on the rotational quantum numbers for both $m$ (here $m$ is corresponding to $-J''$ for P-branch transitions and $J''+1$ for R-branch) and $K$ was considered in the fitting procedure based on the following equation:

$$\gamma(m, K) = a_0 + a_1 \cdot m + a_2 \cdot K + a_3 \cdot m^2 + a_4 \cdot K^2 + a_5 \cdot m \cdot K$$

(6)

The water-vapor broadening Lorentz half widths for ammonia versus the quantum numbers $m$ and $K$ as shown in Fig. 7 help us to visualize the rotational dependence of the broadening coefficients, as they generally decrease with $m$ but increase with $K$. The data reported in (Owen et al., 2013; Sur et al., 2016) were then fitted with a least squares procedure to a polynomial function, namely Eq. (6). The values for these coefficients are $a_0 = 0.52318708$, $a_1 = -0.01249144$, $a_2 = -0.05793872$, $a_3 = -0.002466$, $a_4 = 0.00117773$, $a_5 = 0.00670474$. Therefore, with $|m| \leq 12$ the fitting results based on Eq. (6) were used, and the uncertainty is within 10% (uncertainty code 5 in HITRAN). For $|m| > 12$, a constant value 0.228 corresponding to the average value of transitions with $|m| = 12$ was applied to the whole dataset, as well as for any unassigned lines. Furthermore, the temperature-dependent exponent was set to 0.9 based on the average of the four measured transitions from (Sur et al., 2016).

2.7 H$_2$S

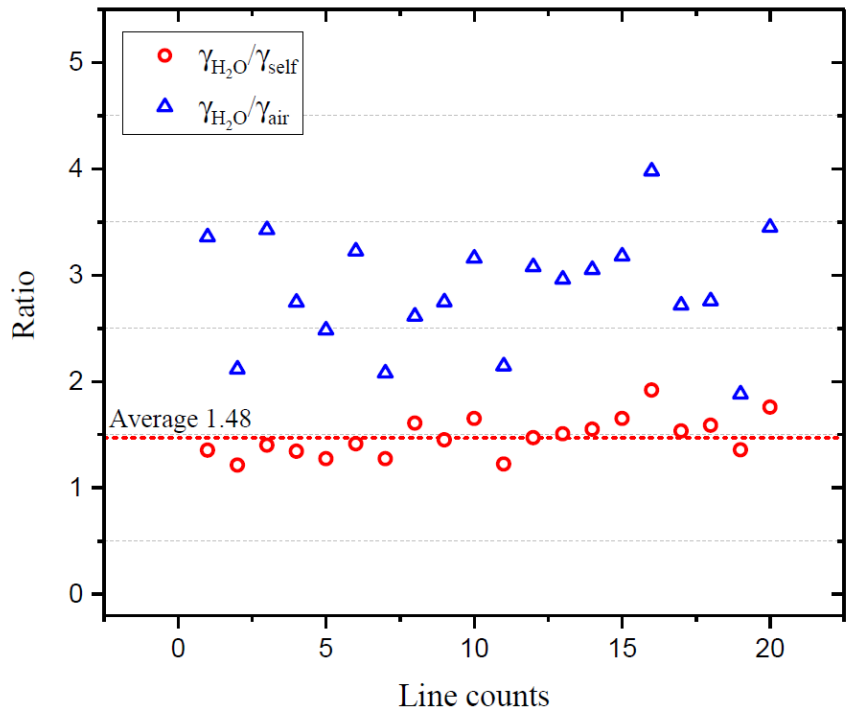

**Figure 8.** The ratio of water-vapor broadening coefficients from calculation (Starikov & Protasevich, 2006) to their corresponding self- and air-broadening values of H$_2$S from HITRAN (Gordon et al., 2017).

While for H$_2$S-H$_2$O, there were no experimental data available in literature, some calculations were available from (Starikov & Protasevich, 2006). The ratio of water-vapor broadening coefficients from calculation to that of self- and air-broadening of H$_2$S is shown in Fig. 8. The derived scaling factor of 1.48 was used to scale the self-broadening coefficients from the HITRAN database. The uncertainty would be larger than 20% (error code 3).

## 3 Working with new data using HITRAN*online* and HAPI

The water broadening parameters and their temperature dependencies described above have been incorporated into the relational structure of the database and are already available via HITRAN*online* (Hill et al., 2016) and HAPI (Kochanov et al., 2016). Both these approaches in acquiring the HITRAN data were described in detail in corresponding papers and in the paper devoted to the HITRAN2016 edition (Gordon et al., 2017).

In HITRAN*online* (https://hitran.org), in order to get the extra line parameters discussed in this paper, the user needs to log in to the system and create a custom output format containing the parameters of interest. The details of this process are described in the dedicated paper by Hill *et al.* (Hill et al., 2016). Fig. 9 shows the example of such output format which includes the water-vapor broadening Lorentz half widths and its temperature dependence exponents, $\gamma_{H2O}$ and $n_{H2O}$ respectively. By clicking on the corresponding checkboxes it is also possible to request their error and reference codes.

**Figure** 9. Sample custom output format creation at HITRAN*online* (https://hitran.org) for getting the water-broadened lines. The line-shape parameters ($\gamma_{H2O}$ and $n_{H2O}$) should be explicitly chosen from the parameter list to extract information relevant to this paper.

In order to calculate absorption coefficients, absorption cross-sections, transmission etc. from the transitions using these foreign-broadening parameters, the HITRAN Application Programming Interface (HAPI) tool is available. It can be obtained at the official web page (https://hitran.org/hapi) and on Github (https://github.com/hitranonline/hapi). HAPI is a set of Python libraries providing means to work with spectroscopic linelists, including calculation of different spectral functions. More details on its usage are given in the documentation which is available at the official HITRAN web page.

Here we give an example which downloads the $CO_2$ linelist in the 1.6 micrometer region accounting for the water broadening using the *fetch* function with explicitly specified "voigt_h2o" parameter group. The *absorptionCrossSection* function was used to calculate absorption cross-sections from the downloaded linelist. The code fragment corresponding directly to the calculation of absorption cross-section is shown in **Fig. 10**. The full code downloading and calculating these sample spectra can be found in the Supplementary materials. In the example given in **Fig. 11**, we calculate and plot absorption cross-sections for $CO_2$ diluted in four different mixtures of air and water. Each mixture is passed to the function through the "Diluent" parameter, which is a dictionary of a type {'air':$VMR_{air}$,'h2o':$VMR_{h2o}$}.

It is worth noting that this approach works for the rest of the foreign broadeners since HAPI and HITRAN*online* follow the similar naming convention for the parameters (see Table 3 of (Kochanov et al., 2016) and Table 3 of the HITRAN2016 paper (Gordon et al., 2017) for broadenings by $CO_2$, He, and $H_2$).

```
16    h.fetch('co2',2,1,6350.0,6375.0,
17            ParameterGroups=('voigt_h2o',))
18
19    mix = [
20            {'air':1.0,'h2o':0.0},
21            {'air':0.7,'h2o':0.3},
22            {'air':0.3,'h2o':0.7},
23            {'air':0.0,'h2o':1.0},
24          ]
25
26    leg = []
27    for d in mix:
28        leg.append(make_legend(d))
29        nu,xsc = h.absorptionCrossSection(
30            SourceTables='co2',
31            Diluent=d,
32            profile='Voigt',
33            WavenumberRange=[NUMIN,NUMAX],
34            )
35        pl.plot(nu,xsc)
```

**Figure** 10. HAPI code snippet for calculating the absorption cross-sections for $CO_2$ lines in 1.6-micrometer spectral range using Voigt line profile and four mixtures of air and water. Full code is available in the Supplementary material.

**4 Conclusions**

In this work the HITRAN database has been extended to include the water-vapor broadening coefficients as well as their temperature dependent exponents for $CO_2$, $O_2$, $CH_4$, CO, $NH_3$, $N_2O$ and $H_2S$ based on semi-empirical models. As shown in Table 3, the different numbers represent the data availability for each molecule in the present work. It is clear that further studies are needed for some of these collisional systems.

Every line for molecules from Table 3 in HITRAN now has the parameters in question. The data can be obtained through HITRAN*online* interface and through HAPI as described in section 4. Procedures described here will allow estimating water-vapor broadening coefficients as well as their temperature-dependence exponents for lines beyond those that are currently provided in the HITRAN database. It is worth noting that HITRAN editions have supplied self-broadening of water-vapor lines. Thus water-vapor broadening of $H_2O$ lines is already provided, although without temperature dependence exponents. Once new measurements and calculations will become available for molecules in Table 3 or other HITRAN molecules, we will consider extending the work presented here.

At the moment there was not enough information to add shifts of spectral lines due to the pressure of water vapor. When a sufficient amount of measurements or calculations of the shifts becomes available, they will be added to the database. With that being said, these shifts will not make substantial differences in the spectral retrievals of the terrestrial atmosphere as shifts are generally from one to two orders of magnitude smaller than the widths (see for instance values in (Delahaye et al., 2016c)) and at 4% abundance will constitute a very small contribution.

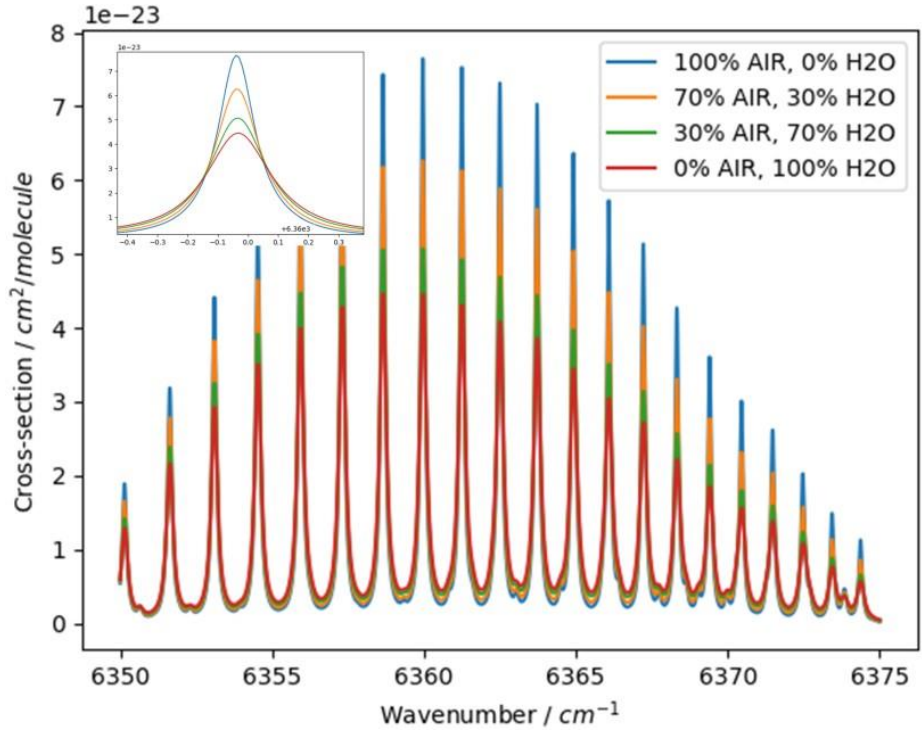

**Figure** 11. Result of the execution of the code give in the **Figure 10**. The strongest line in this region is centered at 6360 cm$^{-1}$ and zoomed in to show appreciable changes in the line shapes for different concentration of the broadeners (see legend).

**Table 3.** Data availability for broadening coefficients (and their half-widths) of spectral lines of different molecules by water vapor.

| Parameter  Molecule | γ | n |
|---|---|---|
| $CO_2$ | 3 | 3 |
| $O_2$ | 3 | 2-3 |
| $CH_4$ | 1-2 | 1 |
| CO | 2-3 | 2-3 |
| $NH_3$ | 1-2 | 0 |
| $N_2O$ | 1 | 0 |
| $H_2S$ | 1 | 0 |

Note: 0 = no data available, 1 = few data available, new HITRAN file contains mostly averages, 2 = some measurements available, allowing semi-classical extrapolations, and 3 = relatively complete set of measurements or calculations available - at least for the room temperature.

**Acknowledgments and Data**

This work is supported by the National Aeronautics and Space Administration AURA (a NASA mission to study Earth's ozone) program grant (NNX17AI78G) and Planetary Data Archiving, Restoration, and Tools (PDART) program grant (NNX16AG51G). The data generated in this paper are available on the HITRAN*online* website: https://hitran.org. Section 3 details how to retrieve and work with that data. Moreover, using procedures described here researchers can estimate broadening coefficients (and their temperature dependencies) for any spectral line of gases in questions even if these lines are not in HITRAN.

feedlot to regional scales. *Journal of Geophysical Research: Atmospheres*, *120*(18), 9718–9738. https://doi.org/10.1002/2015JD023241

Owen, K., & Farooq, A. (2014). A calibration-free ammonia breath sensor using a quantum cascade laser with WMS 2f/1f. *Applied Physics B*, *116*(2), 371–383. https://doi.org/10.1007/s00340-013-5701-1

Owen, K., Es-sebbar, E., & Farooq, A. (2013). Measurements of $NH_3$ linestrengths and collisional broadening coefficients in $N_2$, $O_2$, $CO_2$, and $H_2O$ near 1103.46 cm$^{-1}$. *Journal of Quantitative Spectroscopy and Radiative Transfer*, *121*, 56–68. https://doi.org/10.1016/j.jqsrt.2013.02.001

Oyafuso, F., Payne, V. H., Drouin, B. J., Devi, V. M., Benner, D. C., Sung, K., et al. (2017). High accuracy absorption coefficients for the Orbiting Carbon Observatory-2 (OCO-2) mission: Validation of updated carbon dioxide cross-sections using atmospheric spectra. *Journal of Quantitative Spectroscopy and Radiative Transfer*, *203*, 213–223. https://doi.org/10.1016/j.jqsrt.2017.06.012

Rosenkranz, P. W. (2001). Retrieval of temperature and moisture profiles from AMSU-A and AMSU-B measurements. *IEEE Transactions on Geoscience and Remote Sensing*, *39*(11), 2429–2435. https://doi.org/10.1109/36.964979

Rosenmann, L., Hartmann, J. M., Perrin, M. Y., & Taine, J. (1988). Accurate calculated tabulations of IR and Raman CO2 line broadening by CO2, H2O, N2, O2 in the 300-2400-K temperature range. *Applied Optics*, 0–5. https://doi.org/10.1364/AO.27.003902

Rosenmann, L., Perrin, M. Y., Hartmann, J. M., & Taine, J. (1988). Diode-laser measurements and calculations of $CO_2$-line-broadening by $H_2O$ from 416 to 805 K and by $N_2$ from 296 to 803 K. *Journal of Quantitative Spectroscopy and Radiative Transfer*, *40*(5), 569–576. https://doi.org/10.1016/0022-4073(88)90137-9

Rothman, L. S., Jacquemart, D., Barbe, A., Chris Benner, D., Birk, M., Brown, L. R., et al. (2005). The HITRAN 2004 molecular spectroscopic database. *Journal of Quantitative Spectroscopy and Radiative Transfer*, *96*(2), 139–204. https://doi.org/10.1016/j.jqsrt.2004.10.008

Rothman, L. S., Gordon, I. E., Barber, R. J., Dothe, H., Gamache, R. R., Goldman, A., et al. (2010). HITEMP, the high-temperature molecular spectroscopic database. *Journal of Quantitative Spectroscopy and Radiative Transfer*, *111*(15), 2139–2150. https://doi.org/10.1016/j.jqsrt.2010.05.001

Rothman, L. S., Gordon, I. E., Babikov, Y., Barbe, A., Chris Benner, D., Bernath, P. F., et al. (2013). The HITRAN2012 molecular spectroscopic database. *Journal of Quantitative Spectroscopy and Radiative Transfer*, *130*, 4–50. https://doi.org/10.1016/j.jqsrt.2013.07.002

Schilt, S. (2010). Impact of water vapor on 1.51 μm ammonia absorption features used in trace gas sensing applications. *Applied Physics B*, *100*(2), 349–359. https://doi.org/10.1007/s00340-010-3954-5

Schilt, Stéphane, Thévenaz, L., Niklès, M., Emmenegger, L., & Hüglin, C. (2004). Ammonia monitoring at trace level using photoacoustic spectroscopy in industrial and environmental applications. *Spectrochimica Acta Part A: Molecular and Biomolecular Spectroscopy*, *60*(14), 3259–3268. https://doi.org/10.1016/j.saa.2003.11.032